\documentclass[12pt]{article}
\usepackage{amssymb,amsfonts}
\usepackage{epsf,epsfig}
\begin{document}
\baselineskip 18pt

\title{Quantum mechanics not on manifold}
\author{P.~N.~Bibikov and L.~V.~Prokhorov\\ \it
Sankt-Petersburg State University}

\maketitle

\vskip5mm

\begin{abstract}
The free scalar field is studied on the Y-junction of three semi
infinite axes  which is the simplest example of a non-manifold
space. It is shown that under an assumption that the junction
point can not gain a macroscopic amount of energy and charge the
transmission rules for this system uniquely follow from
conservation of energy and charge. This result is also obtained in
the discrete version of the model. Some alternative approaches to
the problem based on quantum mechanics of Hamiltonian systems with
constrains are discussed.
\end{abstract}

\section {Introduction.}
Hamiltonian mechanics on manifolds now is practically completed
[1,2]. But there appeared serious need in formulation mechanics
not on manifolds. The problems arise both in nanoelectronics and
the string theory. For example, three quantum wires with Y
junction do not compose a manifold (they are not homeomorphic to
some Euclidean space). One can compose of strings a network [3,4]
which also is not a manifold. But it is extremely important to
have the Hamiltonian formalism on such structures. In
nanoelectronics --- to describe motion of electrons, in the string
theory --- the superstring network models the 3-dimensional space,
and one should know how to describe propagation of excitations
over the structure [4].

There is no regular theory of such processes. As a first step to
this end we study a 3-tail system --- a Y junction of three
semi-infinite sets of classical harmonic oscillators and a theory
of free classical scalar field on such a "3-ray star".

Even more serious problems arise when one turns to quantum
description. Quantum mechanics (QM) can be deduced from the
classical one only in the Euclidean space (this Dirac's recipe is
confirmed by experiments). Even the curved spaces causes serious
difficulties. There are two points of view in
this case:\\
1) The curved space is considered as that embedded into the plane
space, and
one has to consider dynamics with constrains.\\
2) QM should be deduced from the classical one without using the
embedding
space.\\
They are two principally different approaches. But neither of them
gives a
unique recipe. In the case 1) there are recipes:\\
(i) The Dirac method (modification of the Poisson brackets) [5].\\
(ii) "The conversion method" [6,7].\\
(iii) "The thin layer method" [8].\\
(iv) "The reduction method" [9,10].\\
The Dirac recipe is not unambiguous [11], the result depends even
on the way one parameterizes the curved space [12]. In the case
(ii) the authors increase the number of unphysical variables. The
approaches (i), (ii) gives different results [11]. In both cases
it is assumed that in QM the unphysical degrees of freedom can not
influence the physical dynamics that is correct only in classical
theory. In the recipe (iii) one approximates the motion on a
surface by motion on a thin layer. This looks reasonable. In the
method (iv) one excludes the normal to the surface motion
demanding
\begin{equation}
\hat P_{\perp}\psi_{ph}=0,
\end{equation}
where $\hat P_{\perp}$ is normal to the surface momentum, and
$\psi_{ph}$ is a state vector from the physical Hilbert space. The
methods (iii) and (iv) give identical results [13] but the latter
allows to avoid rather cumbersome calculations. As for the case 2)
--- QM cannot be deduced unambiguously from the classical one
because there are a lot of "quantum mechanics" giving in the limit
$\hbar\rightarrow 0$ the same classical one.

The situation gets worse if one tries to formulate QM not on
manifold (e.g. on three semi-infinite straight lines having one
general point). In the present paper the classical field is
studied on a "s-ray star". The junction then plays a role of a
potential (scatterer). The corresponding scattering amplitudes are
calculated.

In fact it gives example of both classical and quantum mechanics
in spaces of this type --- scattering of a classical free
relativistic field here is in fact identical to scattering of a
particle in relativistic QM. It turns out that the scattering
amplitudes do not depend on the angle between the rays. There is
no special reason for such an effect because sets of harmonic
oscillators vibrating in the direction orthogonal to the embedding
the "star" plane (i.e. it is supposed that all the rays belong to
the plane) model the dynamics
--- the oscillations do not depend on the angles between the rays.

Importance of this problem for strings is self evident. Gradually
it becomes clear that at the Planck scales matter manifests itself
in form of strings. Polymers and nanostructures are important for
modern technologies. Studying of strings is of special interest
because a 3D-network of superstrings can model the physical space,
thus leading to unification of all interactions, including
gravitation [4,10].

It is worth to note that the 3-tail problem is analogous to the
3-body scattering problem in quantum mechanics.

\section{General properties of $S$-matrix}
Let us consider a complex scalar field $\varphi$ defined on an
Y-junction of three strings with the spatial coordinates
$x\in[0,\infty)$, $y\in[0,\infty)$ and $z\in[0,\infty)$. The
junction point corresponds to $x=y=z=0$. On each string the field
$\varphi$ satisfies the Klein-Fock-Gordon equation (we take
$\hbar=1$, $c=1$),
\begin{equation}
\frac{\partial^2\varphi}{\partial
t^2}=\frac{\partial^2\varphi}{\partial q^2}- m^2\varphi,\quad
q=x,y,z,\quad q>0.
\end{equation}
Our purpose is to obtain a global solution defined on the whole
Y-junction. First of all we demand that the global solution is
continuous at the junction point,
\begin{equation}
\lim_{x\rightarrow0}\varphi(x)=\lim_{y\rightarrow0}\varphi(y)=\lim_{z\rightarrow0}\varphi(z).
\end{equation}
This condition was also postulated in \cite{14} together with the
following one,
\begin{equation}
\varphi_x|_{x=0}+\varphi_y|_{y=0}+\varphi_z|_{z=0}=0,
\end{equation}
where $\varphi_q\equiv\partial_q\varphi$. The latter condition was
used in \cite{14} but its physical sense was not clarified. In the
present paper we show that together with (3) the condition (4)
guarantees both the energy and the charge conservation for our
system.

Local solutions of Eq. (2) on strings satisfy the superposition
principle. It is natural to begin the investigation with study of
a monochromatic wave propagating from $x=\infty$,
\begin{eqnarray}
\varphi(k,x,t)&=&{\rm e}^{-i(\omega t+kx)}+R(k){\rm e}^{-i(\omega t-kx)},\nonumber\\
\varphi(k,y,t)&=&T_x(k){\rm e}^{-i(\omega t-ky)},\quad
\varphi(k,z,t)=T_y(k){\rm e}^{-i(\omega t-kz)}.
\end{eqnarray}
Here $R(k)$ and $T(k)$ are correspondingly the reflection and
transition coefficients, while
\begin{equation}
\omega^2=k^2+m^2,\quad \omega\geq m.
\end{equation}
The incoming particle has momentum $k>0$.

According to (3) $T_x(k)=T_y(k)=1+R(k)$. A {\it unitarity}
condition,
\begin{equation}
|R(k)|^2+2|R(k)+1|^2=1,
\end{equation}
will be proved in the next section. According to it the
coefficient $R(k)$ may be parameterized as follows,
\begin{equation}
R(k)=\frac{1}{3}{\rm e}^{i\theta(k)}-\frac{2}{3}.
\end{equation}

\section{$S$-matrix and conservation of energy and charge}

The Eq. (2) on a line corresponds to the Lagrangian,
\begin{equation}
{\cal L}=\frac{1}{2}(\partial_0\bar\varphi\partial_0\varphi-
\partial_1\bar\varphi\partial_1\varphi-m^2\bar\varphi\varphi),
\end{equation}
where $\partial_0$ and $\partial_1$ denote differentiations with
respect to time and spatial coordinate $q$. The energy-momentum
tensor of the field is given by the general formula \cite{15}
\begin{equation}
T^{ij}=\frac{\partial{\cal
L}}{\partial(\partial_i\varphi)}\partial^j\varphi+\frac{\partial{\cal
L}}{\partial(\partial_i\bar\varphi)}\partial^j\bar\varphi-g^{ij}{\cal
L},
\end{equation}
Here $g^{ij}$ is the Minkowski tensor $g^{ij}=diag(1,-1)$ and the
derivatives $\partial^j$ are related to $\partial_j$ by,
$\partial^i=g^{ij}\partial_j$. Using (9) we obtain,
\begin{equation}
T^{00}=\frac{1}{2}(\partial_0\bar\varphi\partial_0\varphi-
\partial_1\bar\varphi\partial_1\varphi-m^2\bar\varphi\varphi),\quad
T^{10}=-(\partial_1\bar\varphi\partial_0\varphi+\partial_0\bar\varphi\partial_1\varphi).
\end{equation}

The energy-momentum conservation condition is given by the
equation,
\begin{equation}
\partial_iT^{ij}=0.
\end{equation}
According to (12) the energy in a segment $q_1\leq q\leq q_2$,
\begin{equation}
E(q_1,q_2)=\int_{q_1}^{q_2}T^{00}(q)dq,
\end{equation}
satisfies the relation,
\begin{equation}
\frac{dE(q_1,q_2)}{dt}=T^{10}(q_1)-T^{10}(q_2).
\end{equation}

For the system (9) on a line with boundary conditions
$\varphi(\pm\infty)\rightarrow0$ Eq. (14) results to conservation
of the energy $E(-\infty,\infty)=const$. Postulating the energy
conservation for the system on the Y-junction we obtain from (14)
the following condition,
\begin{equation}
T^{10}(x)|_{x\rightarrow0}+T^{10}(y)|_{y\rightarrow0}+T^{10}(z)|_{z\rightarrow0}=0,
\end{equation}
or according to (11),(12),
\begin{equation}
\bar\varphi_t(\varphi_x+\varphi_y+\varphi_z)+(\bar\varphi_x+\bar\varphi_y+\bar\varphi_z)\varphi_t|_{x=y=z=0}=0.
\end{equation}

Though this condition is weaker than (4) it puts a strong enough
restriction on the function $R(k)$. Substituting in (16) the
monochromatic solution (5) we obtain the unitarity condition (7).
However Eq. (16) must be also true for a superposition of several
monochromatic waves with different $k$ or equivalently for the
Fourier sum,
\begin{equation}
\varphi_{in}(x,t)=\sum_ka(k){\rm e}^{-i(\omega_k t-kx)}.
\end{equation}
Substituting (17) into (16) we have to take into account the
interference of exponents with different $\omega_k$. Since the
expression (11) for $T^{10}$ is bilinear with respect to $\varphi$
and $\bar\varphi$ crossing terms originate from {\it two}
monochromatic waves with different frequencies. Therefore, in
order to obtain the corresponding restrictions on the function
$R(k)$ it is sufficient to study the two-mode solution,
\begin{eqnarray}
\varphi(k_1,k_2,x,t)&=&{\rm e}^{-i(\omega_{k_1}
t+k_1x)}+R(k_1){\rm e}^{-i(\omega_{k_1} t-k_1x)}
+{\rm e}^{-i(\omega_{k_2}t+k_2x)}+R(k_2){\rm e}^{-i(\omega_{k_2}t-k_2x)},\nonumber\\
\varphi(k_1,k_2,y,t)&=&(1+R(k_1)){\rm e}^{-i(\omega_{k_1}
t-k_1y)}+(1+R(k_2)){\rm e}^{-i(\omega_{k_2}
t-k_2y)},\nonumber\\
\varphi(k_1,k_2,z,t)&=&(1+R(k_1)){\rm e}^{-i(\omega_{k_1}
t-k_1z)}+(1+R(k_2)){\rm e}^{-i(\omega_{k_2}t-k_2z)}.
\end{eqnarray}
We have suggested here that a wave number does not change after
the scattering.

Substituting (18) into (16) and extracting the constant terms we
obtain Eq. (7). However the terms proportional to ${\rm
e}^{i(\omega_{k_1}-\omega_{k_2})t}$ give the following condition,
\begin{equation}
\omega_{k_1}k_2(1+\bar R(k_1))(1+3R(k_2))+\omega_{k_2}k_1(1+3\bar
R(k_1))(1+R(k_2))=0,
\end{equation}
as well as its complex conjugate. With use (6) and (8) this two
relations give,
\begin{equation}
{\rm
e}^{i\theta(k)}=\frac{k+i\alpha\sqrt{k^2+m^2}}{k-i\alpha\sqrt{k^2+m^2}}.
\end{equation}
Here $\alpha$ is a real constant.

Another important conserving quantity is charge \cite{15} related
to the current,
\begin{equation}
j_{\mu}=i(\bar\varphi\partial_{\mu}\varphi-\varphi\partial_{\mu}\bar\varphi),\quad
\mu=0,1.
\end{equation}
From Eq. (2) it follows that,
\begin{equation}
\partial_0j_0+\partial_1j_1=0,
\end{equation}
and analogously to (15) postulating charge conservation we obtain
the following {\it additional} transmission condition,
\begin{equation}
j_1(x)|_{x\rightarrow0}+j_1(y)|_{y\rightarrow0}+j_1(z)|_{z\rightarrow0}=0,
\end{equation}
or
\begin{equation}
\bar\varphi(\varphi_x+\varphi_y+\varphi_z)-(\bar\varphi_x+\bar\varphi_y+\bar\varphi_z)\varphi=0.
\end{equation}
Again we have to check it substituting the solutions (5) and (18).
The substitution of (5) into (24) gives again the unitarity
condition (7), however the substitution of (18) results to,
\begin{equation}
k_2(1+\bar R(k_1))(1+3R(k_2))+k_1(1+3\bar R(k_1))(1+R(k_2))=0,
\end{equation}
or
\begin{equation}
{\rm e}^{i\theta(k)}=\frac{k+i\beta}{k-i\beta},
\end{equation}
where $\beta$ is a new real constant.

As we see from (20) and (26) the energy and charge will conserve
simultaneously only for,
\begin{equation}
\alpha=\beta=0,
\end{equation}
or,
\begin{equation}
\alpha=\beta=\infty.
\end{equation}
In the first case,
\begin{equation}
R(k)=-\frac{1}{3},\quad T(k)=\frac{2}{3},
\end{equation}
however in the second one,
\begin{equation}
R(k)=-1,\quad T(k)=0.
\end{equation}
For $T(k)=0$ the three strings behaves as disjoint ones, so the
solution (30) is of little physical interest. On the other hand
substituting (5) into (4) we find that for the monocromatic waves
the conditions (4) and (29) are equivalent. Since both of them are
linear this equivalence is also true for a general solution (17).
An outstanding feature of the solution (29) is universality. It
does not depend on $k$. This is important for modeling of space by
a network composed by strings \cite{4},\cite{10}.

We conclude that Eq. (4) represents the only nontrivial condition
compatible with continuity condition (3), superposition principle
and both the energy and charge conservation.

\section{Harmonic oscillators network approximation}

It is instructive to approximate our system by a harmonic network.
The latter consists of three linear chains of harmonic oscillators
related to variables $\varphi_{q,n}$, where $n=1,2,...$ and
$q=x,y,z$ added by the junction point oscillator described by
$\varphi_0$. The Lagrangian is given by,
\begin{eqnarray}
L&=&\frac{1}{2}\sum_q\sum_n\Big[\dot\varphi_{q,n}^2-\frac{1}{\Delta^2}(\varphi_{q,n+1}-\varphi_{q,n})^2-
m^2\varphi_{q,n}^2\Big]\nonumber\\
&+&\frac{1}{2}\Big[\dot\varphi_0^2-\frac{1}{\Delta^2}\sum_q(\varphi_0-\varphi_{q,1})^2-m^2\varphi_0^2\Big],
\end{eqnarray}
where $\Delta$ is the lattice constant.

The Lagrangian (31) gives the following equations of motion:
\begin{eqnarray}
\ddot\varphi_0&=&\frac{1}{\Delta^2}(\sum_q\varphi_{q,1}-3\varphi_0)-m^2\varphi_0,\\
\ddot\varphi_{q,1}&=&\frac{1}{\Delta^2}(\varphi_{q,2}+\varphi_0-2\varphi_{q,1})-
m^2\varphi_{q,1},\\
\ddot\varphi_{q,n}&=&\frac{1}{\Delta^2}(\varphi_{q,n+1}+\varphi_{q,n-1}-2
\varphi_{q,n})-m^2\varphi_{q,n},\quad n>1.
\end{eqnarray}

The following solution,
\begin{eqnarray}
\varphi_{x,n}(t)&=&{\rm e}^{-i(\omega_kt+kn)}+R(k){\rm
e}^{-i(\omega_kt-k\Delta n)},\nonumber\\
\varphi_{y,n}(t)=\varphi_{z,n}(t)&=&(R(k)+1){\rm e}^{-i(\omega_kt-k\Delta n)},\nonumber\\
\varphi_0(t)&=&(R(k)+1){\rm e}^{-i\omega_kt}.
\end{eqnarray}
is a discrete analog of (5). The normal frequencies,
\begin{equation}
\omega^2_k=\frac{4}{\Delta^2}\sin^2\frac{k\Delta}{2}+m^2,
\end{equation}
in the limit $\Delta\rightarrow0$ coincide with (6).

Substituting (35) into (32) and taking into account the condition,
\begin{equation}
\ddot\varphi_0+m^2\varphi_0=-\frac{4}{\Delta^2}\sin^2\frac{k\Delta}{2}(1+R(k)){\rm
e}^{-i\omega_kt},
\end{equation}
(see (36)) we obtain,
\begin{equation}
[4\sin^2\frac{k\Delta}{2}+3({\rm
e}^{ik\Delta}-1)](R(k)+1)=2i\sin{k\Delta}.
\end{equation}
Since ${\rm
e}^{ik\Delta}-1=2i\sin{\frac{k\Delta}{2}}\cos{\frac{k\Delta}{2}}-2\sin^2{\frac{k\Delta}{2}}$
and
$\sin{k\Delta}=2\sin{\frac{k\Delta}{2}}\cos{\frac{k\Delta}{2}}$,
we reduce (38) to the following form,
\begin{equation}
(3i\cos{\frac{k\Delta}{2}}-\sin{\frac{k\Delta}{2}})(R(k)+1)=2i\cos{\frac{k\Delta}{2}}
\end{equation}
or
\begin{equation}
{\rm
e}^{i\theta(k)}=-\frac{\sin\frac{k\Delta}{2}+3i\cos\frac{k\Delta}{2}}
{\sin\frac{k\Delta}{2}-3i\cos\frac{k\Delta}{2}}.
\end{equation}
In the limit $\Delta\rightarrow0$ using Eq. (8) we obtain for
$R(k)$ and $T(k)$ the expression (29).

The authors are grateful to B.~S.~Pavlov for the interesting and
helpful discussion.

\end{document}